\documentclass[12pt,preprint]{aastex}

\slugcomment{\today}

\shorttitle{Cosmic Ray Diffusion}
\shortauthors{Ryu et al.}

\begin{document}

\title{The Effect of Cosmic Ray Diffusion on the Parker Instability}

\author{Dongsu Ryu\altaffilmark{1},
        Jongsoo Kim\altaffilmark{2},
        Seung Soo Hong\altaffilmark{3}
        and T. W. Jones\altaffilmark{4}}

\altaffiltext{1}
{Department of Astronomy \& Space Science, Chungnam National University,
Daejeon 305-764, Korea; ryu@canopus.chungnam.ac.kr}
\altaffiltext{2}
{Korea Astronomy Observatory, 61-1, Hwaam-Dong, Yusong-Ku, Taejon 305-348,
Korea; jskim@kao.re.kr}
\altaffiltext{3}
{Astronomy Program, School of Earth and Environmental Sciences, Seoul
National University, Seoul 151-742, Korea; sshong@astro.snu.ac.kr}
\altaffiltext{4}
{Department of Astronomy, University of Minnesota, Minneapolis, MN 55455;
twj@msi.umn.edu}

\begin{abstract}
The Parker instability, which has been considered as a process
governing the structure of the interstellar medium, is induced by
the buoyancy of magnetic field and cosmic rays. In previous studies,
while the magnetic field has been fully incorporated in the context
of isothermal magnetohydrodynamics, cosmic rays have been normally
treated with the simplifying assumption of infinite diffusion along
magnetic field lines but no diffusion across them. The cosmic
ray diffusion is, however, finite. In this work, we take into account
fully the diffusion process of cosmic rays in a linear
stability analysis of the Parker instability. Cosmic rays are described
with the diffusion-convection equation. With realistic values of cosmic
ray diffusion coefficients expected in the interstellar medium, we show
that the result of previous studies with the simplifying assumption on
cosmic ray diffusion applies well. Finiteness of parallel diffusion
decreases the growth rate of the Parker instability, while the relatively
smaller perpendicular diffusion has no significant effect. We discuss
the implication of our result on the role of the Parker instability
in the interstellar medium.

\end{abstract}

\keywords{cosmic rays --- instabilities --- ISM: magnetic fields --- MHD }

\section{Introduction}

In stability analyses of the interstellar medium (ISM), \citet{par66,par67}
put forward a simple model of the ISM which is composed of a single phase gas,
magnetic field and cosmic rays (CRs) under external, uniform, vertical
gravity. It was assumed that gas pressure originates from
the ram motion of cloudlets, rather than the thermal motion of atoms or
molecules, so the velocity dispersion of cloudlets was taken as the
sound speed. In equilibrium, the magnetic field has only a regular component
and the ratios of pressures of magnetic field and CRs to gas pressure
are constant. In addition, the CR dynamics was simplified by setting
$d\delta P_c/dt=0$, based on the assumption that the CR pressure is uniform
along magnetic field lines\footnote{Later, \citet{shu74} refined this
treatment in a more intuitive form that the diffusion along magnetic
field lines is very large but the diffusion across field lines is
negligible. Both formulations of \citet{par66,par67} and \citet{shu74}
turn out to be same in the linear regime. See \S 3.2.}. Then, he showed that
the equilibrium state is subject to an instability, which is now known
as the Parker instability.

The Parker's work has since been elaborated upon. For instance,
\citet{giz93}, \citet{kim97}, and \citet{kim98} investigated the
modification of the Parker instability under nonuniform gravities.
It was found that the linear growth rate increases under the gravities
described by linear and hyperbolic-tangent functions. \citet{kim00}
and \citet{san00} incorporated the multi-component nature of
the ISM in a realistic gravity model, where the growth
timescale turns out to be $\sim3\times10^7$ years and the length scale
enlarges up to $\sim3$ kpc. The effect of the irregular, random component
of magnetic field was studied in \citet{par00} and \citet{kim01}.
With the strength of the random component comparable to the regular one
\citep[see, e.g.,][]{bbmss96,zh97}, it was shown that the Parker instability
can be completely stabilized.

CRs form an important constituent of the ISM, with their energy density
comparable to those of gas and magnetic field \citep[see, e.g.,][]{be87}.
The analyses of \citet{par66,par67} and \citet{shu74} showed that CRs
play a significant role in the development of the Parker instability
by widening the range of unstable wavelengths and increasing the growth
rate under the limit of $\kappa_{\parallel}\rightarrow\infty$ (very large
diffusion along magnetic field lines) and $\kappa_{\perp}=0$ (negligible
diffusion across field lines). However, it is certainly true that
$\kappa_{\parallel}$ and $\kappa_{\perp}$ are finite (see \S 2.2
for details), but there has been no follow-up work on the effect of
CRs with finite diffusion in the Parker instability. In an approach based
on a different perspective, \citet{nel85} incorporated the dynamics of
CRs by approximating their pressure as
\begin{equation}
\label{eq:anisopcr}
P_{{\rm c},ij} = P_{{\rm c},\perp} \delta_{ij} -
(P_{{\rm c},\perp}-P_{{\rm c},\parallel}) \frac{B_iB_j}{B^2}
\end{equation}
(for comparison, see equation~(\ref{eq:kappa}) for the spatial diffusion
tensor). Although the diffusion process was not included, the anisotropic
nature of CR dynamics was taken into account. The surprising result was that
the anisotropic CR pressure works towards  stabilizing
the instability, contrary to the common belief that CRs act as
one of the agents to induce the instability itself.

In this paper, we describe an linear analysis analysis where CR dynamics
is incorporated with the diffusion-convection equation
\citep[see, e.g.,][]{ski75}, and their effect on the Parker instability
is addressed. Previously, \citet{kp83} attempted a similar analysis.
They argued that CRs enhance the instability, by showing that the critical
value of the gas adiabatic index for the instability increases due CRs.
Here, we first estimate the values of $\kappa_{\parallel}$ and
$\kappa_{\perp}$ which are applicable in the analysis, and then derive
the dispersion relation. We show that the growth rate and the range
of unstable wavelengths increase due to CRs. Our result confirms
the validity of the analyses of
\citet{par66} and \citet{shu74} at a quantitative level, but
disagrees with that of \citet{nel85}. Recently, \citet{han00} studied
the Parker instability triggered by the CRs injected in supernova remnants.
They solved numerically the flux tube equation for the magnetic field along
with the diffusion-convection equation for CRs. Their work is the first
example which took into account the CR diffusion in the Parker instability.
However, it still needs to be quantified how much CR diffusion affects
the range of unstable wavelengths and the growth rate. This paper addresses
that specific issue.

In \S 2 the stability analysis is described and the dispersion relation
is driven. Also discussion on the CR diffusion tensor is presented.
Interpretation of the dispersion relation is described in \S 3. Summary
and discussion on the implications of our result are given in the final
section.

\section{LINEAR STABILITY ANALYSIS}

\subsection{Basic Equations}

The equation set for our purpose is the combination of the
MHD equations and the CR diffusion-convection equation,
\begin{equation}
\label{eq:continuity}
\frac{\partial \rho}{\partial t} + \nabla \cdot (\rho v) = 0,
\end{equation}
\begin{equation}
\label{eq:momentum}
\rho\left[ \frac{\partial v}{\partial t} + (v \cdot \nabla)v \right]
= - \nabla \left( P_{\rm g} + P_{\rm c} + \frac{B^2}{8\pi} \right)
  + \frac{1}{4\pi} B \cdot \nabla B + \rho g,
\end{equation}
\begin{equation}
\label{eq:induction}
\frac{\partial B}{\partial t} = \nabla \times (v \times B),
\end{equation}
\begin{equation}
\label{eq:gas_energy}
\frac{\partial P_{\rm g}}{\partial t}+v \cdot \nabla P_{\rm g} + 
\gamma_{\rm g} P_{\rm g} \nabla \cdot v = 0,
\end{equation}
\begin{equation}
\label{eq:cr_energy}
\frac{\partial P_{\rm c}}{\partial t}+v \cdot \nabla P_{\rm c} 
+ \gamma_{\rm c} P_{\rm c} \nabla \cdot v
 = \nabla (\left<\kappa_{ij}\right> \nabla P_{\rm c}) + S_0,
\end{equation}
where the subscripts g and c stand for gas and CRs. CRs are described
by the two-fluid model, which is derived from the second particle
momentum moment of the well-known CR Fokker-Planck equation \citep{ski75}.
Hence, CRs are described in equation (\ref{eq:cr_energy}) by a pressure, 
plus an equation of state for the CRs represented by the adiabatic index, 
$\gamma_{\rm c} = 1+{P_{\rm c}}/{E_{\rm c}}$, instead of a full momentum
distribution function \citep[see][for details of the two
fluid model]{dru81,jon90}. $\left<\kappa_{ij}\right>$ is the energy
weighted mean diffusion tensor of CRs (see \S 2.2 for discussion).
The source term, $S_0$, in the CR pressure equation
is introduced to set up an initial equilibrium state (see \S 2.3), not to
describe the injection from thermal particles to CR particles. We
ignore that process, along with shock acceleration of CRs \citep[see, e.g.,]
[for details of shock acceleration]{dru83,be87}. There are no shocks
in the regime of linear stability analyses.

As pointed out by \citet{par66,par67}, on the scale where
the Parker instability is relevant, the dominant contribution
to gas pressure would not come from
the thermal motions of atoms or molecules, but would come from the turbulent
motions of cloudlets. Then, the value of the ``effective'' adiabatic index
for gas, $\gamma_{\rm g}$, should be determined by considering the detailed
mechanisms involved, such as supernova explosions, stellar winds, the
Galactic differential rotation, cloud-cloud collisions, turbulence
dissipation, and etc. Although there has been much progress in the studies
of each mechanism, the determination of $\gamma_{\rm g}$ for an ensemble of
cloudlets is less well understood. Hence, here we simply set
$\gamma_{\rm g}=1$, assuming that the cloudlet random motions are constant.

The adiabatic index of the CRs,
$\gamma_{\rm c} = 1+{P_{\rm c}}/{E_{\rm c}}$, can be simply related to the
form of the CR momentum distribution if the latter is a power law
with an index between 4 and 5.
In particular a momentum distribution
\begin{equation}
\label{eq:crdist}
f(q) \propto p^{-q}~~~{\rm with}~~~q \simeq 14/3
\end{equation}
appropriate for Galactic CRs \citep[see, e.g.,][]{be87},
leads to ${P_{\rm c}}/{E_{\rm c}} = (q-3)/3 \simeq 5/9$.
Hence,
$\gamma_{\rm c} = 14/9$ is used in our analysis.

\subsection{Cosmic Ray Diffusion Tensor}

The frequently used form of the CR diffusion tensor is
\begin{equation}
\label{eq:kappa}
\kappa_{ij} = \kappa_{\perp} \delta_{ij} -
(\kappa_{\perp}-\kappa_{\parallel}) \frac{B_iB_j}{B^2}
+\epsilon_{ijk} \kappa_A \frac{B_k}{B},
\end{equation}
where $B_i$ is the magnetic field vector; $\kappa_{\parallel}$
and $\kappa_{\perp}$ are the diffusion coefficients along and across
mean field, respectively, and $\kappa_A$ represents the curvature and
gradient drifts
\citep[see, e.g.,][for discussions on $\kappa_{ij}$]{bm97,gia99,clp02}.

\citet{gia99} and \citet{clp02} used Monte Carlo simulations in modeling
turbulent magnetic fields and estimated the diffusion properties.
With the energy ratio of the random to total magnetic fields
\begin{equation}
\label{eq:chi}
\chi = \frac{\delta B^2}{B_0^2+\delta B^2},
\end{equation}
\citet{clp02} showed that whenever $\chi<1$, Bohm diffusion 
($\kappa \propto p/\sqrt{p^2 + m^2c^2}$) does not
apply, but the quasi-linear approximation does for parallel diffusion.
They found that
\begin{equation}
\label{eq:kapp_pall}
\kappa_{\parallel} = \frac{v r_L}{3h}~~~{\rm with}~~~
h \simeq 0.4 \chi (r_L k_{min})^{2/3},
\end{equation}
for the Kolmogorov turbulence. Here, $r_L$ and $k_{min}$ represent
the Larmor radius and the minimum wavenumber of the Kolmogorov spectrum,
respectively. The formal quasi-linear result would replace 0.4 by $\pi/6$.
Note that equation~(\ref{eq:kapp_pall}) applies even when $\chi=0.99$.

Based on the GALPROP model of CR propagation in the ISM, \citet{sm98} found
that $\kappa_{\parallel} \simeq 6\times10^{28} {\rm cm}^2/{\rm sec}$ at
the rigidity of $r_L/B_0 = 3 {\rm GV}$ using isotropic diffusion.
Matching this with equation~(\ref{eq:kapp_pall}), $\kappa_{\parallel}$
can be estimated. Taking the momentum distribution in
equation~(\ref{eq:crdist}) and letting the energy weighted mean diffusion be
\begin{equation}
\label{eq:meankappa}
\left<\kappa\right> = \frac{\int\kappa\left(\sqrt{p^2+1}-1\right)
f(q)p^2dp}{\int\left(\sqrt{p^2+1}-1\right)f(q)p^2dp},
\end{equation}
we get
\begin{equation}
\label{eq:meankappa_pall}
\left<\kappa_{\parallel}\right> \simeq 2.5\times10^{28}\left(\frac{0.2}{\chi}
\right) \left(\frac{A}{Z}\right)^{\frac{1}{3}}\left(\frac{3\mu{\rm G}}{B_0}
\right)^{\frac{1}{3}}\left(\frac{L_{max}}{200{\rm pc}}\right)^{\frac{2}{3}},
\end{equation}
where $A$ and $Z$ are the CR atomic number and charge, respectively, and
$L_{max}$ is the coherent length of the regular component of magnetic
field. Note that equation~(\ref{eq:meankappa_pall}) is derived with
$\chi\simeq0.2$, but the result of \citet{sm98} was based on isotropic
diffusion ($\chi = 1$). However, the normalization seems to be uncertain,
at least, by a factor of a few anyway, since the details of CR propagation
are not really understood.

\citet{clp02} found the perpendicular diffusion to behave according to
\begin{equation}
\label{eq:kappa_perp}
\kappa_{\perp} \simeq 0.2 \chi^{2/3} \kappa_{\parallel},
\end{equation}
rather than $\kappa_{\perp} \sim 10^{-6}\kappa_{\parallel}$
which is predicted in the quasi-linear result. Also, from the argument
based on the escape of the Galactic CRs and their lifetime, \citet{gia99}
drew a consistent value for the perpendicular diffusion,
$\kappa_{\perp} \simeq 0.02 - 0.04 \kappa_{\parallel}$.
On the other hand, \citet{bm97} argued that 
\begin{equation}
\label{eq:kappa_A}
\kappa_A \simeq  \left(\frac{c}{r_L}\tau_{\rm decorr}\right) \kappa_{\perp}
~~~{\rm with}~~~\frac{c}{r_L}\tau_{\rm decorr} \sim 2 \frac{r_L}{L_{max}}
\frac{1}{\chi},
\end{equation}
where $\tau_{\rm decorr}$ is the CR decorrelation time. So with
$r_L/L_{max} \ll 1$, it is expected that $\kappa_A \ll \kappa_{\perp}$
and $\kappa_A$ can be neglected to a first approximation.

In the rest of the paper, brackets will be dropped in the mean diffusion
coefficients, for simplicity.

\subsection{Initial Equilibrium State}

A stability analysis is started by setting up the initial equilibrium
configuration. We employ the one originally suggested by \citet{par66,par67}.
In the Cartesian coordinates $(x,y,z)$, the azimuthal magnetic field is
set to lie along the $y$-direction $(0,B_0[z],0)$, and the externally given
uniform gravity to accelerate in the negative $z$-direction $(0,0,-g)$. Then,
the initial state of mass density, $\rho_0$, gas pressure, $P_{\rm g0}$,
CR pressure, $P_{\rm c0}$, and magnetic field, $B_0$, are described by an
exponential function
\begin{equation}
\label{eq:istate}
\frac{\rho_0(z)}{\rho_0(0)} =
\frac{P_{\rm g0}(z)}{P_{\rm g0}(0)} =
\frac{P_{\rm c0}(z)}{P_{\rm c0}(0)} =
\frac{B_0^2(z)}{B_0^2(0)} =
\exp \left( - \frac{z}{H} \right),
\end{equation}
where $H=(1+\alpha+\beta)a^2/g$ and $a$ is the isothermal sound speed
(with $\gamma_{\rm g}=1$). The scale height ($160$ pc) and the velocity
dispersion ($6.4$ km s$^{-1}$) of interstellar clouds will be used for
$H$ and $a$ \citep[see, e.g.,][]{fl73}. $\alpha$ is the ratio of initial
magnetic to gas pressures and $\beta$ is the ratio of initial CR to gas
pressures, respectively, and they are assumed to be constant.

Special attention needs to be given to the initial equilibrium of
equation~(\ref{eq:cr_energy}). Non-zero $\kappa_{\perp}$ would cause CRs
to diffuse upwards. The source term,
$S_0 = -\kappa_{\perp} P_{\rm c0}(z) / H^2$, was included to balance it.
This ad hoc treatment, however, would introduce spurious features in
the stability properties, and the interpretation of it should be done
with caution (see \S 3.3 and 3.4). Of course, with $\kappa_{\perp}=0$,
this problem does not appear.

\subsection{Linearized Perturbation Equations}

Here, we focus as in \citet{par66} on the stability in the $(y,z)$ plane
defined by the initial magnetic field and gravity.
The analysis becomes simplified,
if the linearization of equations~(\ref{eq:continuity})-(\ref{eq:cr_energy})
is proceeded with dimensionless quantities \citep{shu74}. With $H$ and
$H/a$ as the normalization units of length and time, we define
dimensionless coordinates and time,
\begin{equation}
y'=y/H; \;\;\; z'=z/H; \;\;\; t'=at/H,
\end{equation}
and introduce dimensionless perturbations of density,
$s$, velocity, ${\bf u}$,
magnetic field, ${\bf b}$, gas pressure, $p_{\rm g}$, and CR pressure,
$p_{\rm c}$. Then, the perturbed state can be written as
\begin{equation}
\label{eq:pstate}
\rho = \rho_0(z)(1+s); \;\;
{\bf v}=a{\bf u}; \;\;
{\bf B}=B_0(z)(\hat{e}_y+{\bf b}); \;\;
P_{\rm g} = P_{\rm g0}(z)(1+p_{\rm g}); \;\;
P_{\rm c} = P_{\rm c0}(z)(1+p_{\rm c}),
\end{equation}
where $\hat{e}_y$ is the unit vector along the $y$-direction. For simplicity,
primes have been dropped in equation~(\ref{eq:pstate}) and will be in
the rest of the paper.

Substituting equation~(\ref{eq:pstate}) into
equations~(\ref{eq:continuity})-(\ref{eq:cr_energy}) and keeping terms only
up to the linear order of perturbations, the linearized perturbation
equations are written as follows:
\begin{equation}
\label{eq:p_continuity}
\frac{\partial s}{\partial t} - u_z +
\frac{\partial u_y}{\partial y}+\frac{\partial u_z}{\partial z} = 0,
\end{equation}
\begin{equation}
\frac{\partial u_y}{\partial t} +
\frac{\partial p_{\rm g}}{\partial y} +
\beta \frac{\partial p_{\rm c}}{\partial y} +
\alpha b_z = 0,
\end{equation}
\begin{equation}
\frac{\partial u_z}{\partial t} + (1+\alpha+\beta)s -
(p_{\rm g}+\beta p_{\rm c}+2\alpha b_y) +
\frac{\partial}{\partial z}(p_{\rm g}+\beta p_{\rm c}+2\alpha b_y) -
2\alpha \frac{\partial b_z}{\partial y} = 0,
\end{equation}
\begin{equation}
\frac{\partial b_y}{\partial t} - \frac{1}{2}u_z +
\frac{\partial u_z}{\partial z} = 0,
\end{equation}
\begin{equation}
\frac{\partial b_z}{\partial t} - \frac{\partial u_z}{\partial y} =0,
\end{equation}
\begin{equation}
\frac{\partial p_{\rm g}}{\partial t} - u_z +
\gamma_{\rm g} \left( \frac{\partial u_y}{\partial y} +
              \frac{\partial u_z}{\partial z} \right) = 0,
\end{equation}
\begin{equation}
\label{eq:p_crenergy}
\frac{\partial p_{\rm c}}{\partial t} - u_z +
\gamma_{\rm c} \left( \frac{\partial u_y}{\partial y} +
              \frac{\partial u_z}{\partial z} \right) -
\kappa_{\parallel} \frac{\partial^2 p_{\rm c}}{\partial y^2} -
\kappa_{\perp} \left( p_{\rm c}-
                      2\frac{\partial p_{\rm c}}{\partial z} +
                       \frac{\partial^2 p_{\rm c}}{\partial z^2} \right) -
(\kappa_{\perp}-\kappa_{\parallel})\frac{\partial b_z}{\partial y} = 0.
\end{equation}
Here, again primes have been dropped in normalized $\kappa_{\parallel}$
and $\kappa_{\perp}$ for simplicity.

\subsection{Dispersion Relation}

The normal mode of perturbations has the following form
\begin{equation}
\label{eq:perturbation}
\left[ \begin{array}{ccccccc}
        s(y,z,t)   \\
        u_y(y,z,t) \\
        u_z(y,z,t) \\
        b_y(y,z,t) \\
        b_z(y,z,t) \\
        p_{\rm g}(y,z,t)  \\
        p_{\rm c}(y,z,t)  \\
       \end{array}
\right] =
\left[ \begin{array}{ccccccc}
        s         \\
        u_y       \\
        u_z       \\
        b_y       \\
        b_z       \\
        p_{\rm g} \\
        p_{\rm c}
       \end{array}
\right]
\exp(nt) \exp(-i \eta y) \exp \left(\frac{1}{2}z-i\zeta z\right),
\end{equation}
where $n$ is the dimensionless growth rate, $\eta$ and $\zeta$ are the
dimensionless wavenumbers along the azimuthal ($y$) and vertical ($z$)
directions, respectively.
The $\exp(z/2)$ factor was included due to the stratified background.
The same notations were used for perturbations themselves in the left side
and their amplitudes  in the right side, because no confusion arises
in later algebra. Substituting equation~(\ref{eq:perturbation}) into
equations~(\ref{eq:p_continuity})-({\ref{eq:p_crenergy}) results in
the following set of equations:
\begin{equation}
\label{eq:s}
ns - i\eta u_y - \left( \frac{1}{2}+i\zeta \right) u_z = 0,
\end{equation}
\begin{equation}
\label{eq:uy}
nu_y - i\eta p_{\rm g} - i\eta\beta p_{\rm c} + \alpha b_z = 0,
\end{equation}
\begin{equation}
\label{eq:uz}
n u_z + (1+\alpha+\beta) s -
\left(\frac{1}{2}+i\zeta\right)(p_{\rm g}+\beta p_{\rm c} + 2\alpha b_y) +
i\eta 2\alpha b_z = 0,
\end{equation}
\begin{equation}
\label{eq:by}
n b_y - i\zeta u_z = 0,
\end{equation}
\begin{equation}
\label{eq:bz}
n b_z + i\eta u_z = 0,
\end{equation}
\begin{equation}
\label{eq:pg}
n p_{\rm g} - i\eta \gamma_{\rm g} u_y -
\left[ 1 - \gamma_{\rm g} \left(\frac{1}{2}-i\zeta\right) \right] u_z = 0,
\end{equation}
\begin{equation}
\label{eq:pc}
\left[n+\kappa_{\parallel}\eta^2 -
      \kappa_{\perp}\left(\frac{1}{2}+i\zeta\right)^2 \right] p_{\rm c} -
i\eta \gamma_{\rm c} u_y -
\left[1-\gamma_{\rm c}\left(\frac{1}{2}-i\zeta\right)\right] u_z +
(\kappa_{\perp}-\kappa_{\parallel}) i\eta b_z = 0.
\end{equation}

The dispersion relation is derived by combining the above seven equations.
Although straightforward, it entails tedious algebra. Here, we
present a few intermediate steps. First, by eliminating $b_z$ in
equation~(\ref{eq:pc}) with equation~({\ref{eq:bz}), we have
\begin{equation}
\label{eq:npc}
\left[n+\kappa_{\parallel}\eta^2 -
        \kappa_{\perp}\left(\frac{1}{2}+i\zeta\right)^2 \right] n p_{\rm c} -
i\eta \gamma_{\rm c} n u_y -
\left\{ \left[1-\gamma_{\rm c}\left(\frac{1}{2}-i\zeta\right)\right]n -
(\kappa_{\perp}-\kappa_{\parallel})\eta^2 \right\} u_z = 0,
\end{equation}
which expresses $p_{\rm c}$ in terms of $u_y$ and $u_z$. Then, upon
substituting $s$ in equation~(\ref{eq:s}), $b_y$ in equation~(\ref{eq:by}),
$b_z$ in equation~(\ref{eq:bz}), and $p_{\rm c}$ in equation~(\ref{eq:npc})
into equations~(\ref{eq:uy}) and (\ref{eq:uz}), we obtain two equations for
$u_y$ and $u_z$
\begin{eqnarray}
\label{eq:dispa}
&&\left\{(n^2+\gamma_{\rm g} \eta^2)
       \left[n+\kappa_{\parallel}\eta^2-
               \kappa_{\perp}\left(\frac{1}{2}+i\zeta\right)^2\right]
       +\beta\gamma_{\rm c}\eta^2 n \right\} u_y
-i\eta \left\{\left[n+\kappa_{\parallel}\eta^2
              -\kappa_{\perp}\left(\frac{1}{2}+i\zeta\right)^2 \right] \right.
\nonumber \\
&& \left.
\times\left[1+\alpha-\gamma_{\rm g}\left(\frac{1}{2}-i\zeta\right)\right]
              +\beta n \left[1-\gamma_{\rm c}\left(\frac{1}{2}-i\zeta
                       \right)\right]
              - (\kappa_{\perp}-\kappa_{\parallel})\beta\eta^2
\right\} u_z = 0,
\end{eqnarray}
and
\begin{eqnarray}
\label{eq:dispb}
&&i\eta \left\{ \left[n+\kappa_{\parallel}\eta^2
                     -\kappa_{\perp}\left(\frac{1}{2}+i\zeta\right)^2\right]
              \left[1+\alpha+\beta-\gamma_{\rm g}\left(\frac{1}{2}+i\zeta
                        \right)\right]
              -\beta\gamma_{\rm c} n\left(\frac{1}{2}+i\zeta \right)
                        \right\} u_y
\nonumber \\
&&
+\Bigg\{\left[n+\kappa_{\parallel}\eta^2
       -\kappa_{\perp}\left(\frac{1}{2}+i\zeta\right)^2 \right]
\left[n^2+2\alpha\eta^2+\beta\left(\frac{1}{2}+i\zeta\right) +
(2\alpha+\gamma_{\rm g})\left(\frac{1}{4}+\zeta^2\right)\right]
\nonumber \\
&&
+\beta\left(\frac{1}{2}+i\zeta\right)
\left\{(\kappa_{\perp}-\kappa_{\parallel})\eta^2
-\left[1-\gamma_{\rm c}\left(\frac{1}{2}-i\zeta\right)\right] n \right\}
\Bigg\}u_z=0.
\end{eqnarray}
Finally, by combining equations~(\ref{eq:dispa}) and (\ref{eq:dispb}),
we get the dispersion relation, which is a 6th order polynomial of $n$
\begin{equation}
\label{eq:polynomial}
n^6 + C_5 n^5 + C_4 n^4 + C_3 n^3 + C_2 n^2 + C_1 n + C_0 = 0,
\end{equation}
where
\begin{equation}
\label{eq:c5}
C_5 = 2 \left[ \kappa_{\parallel} \eta^2 -
               \kappa_{\perp}\left(\frac{1}{2}+i\zeta\right)^2 \right],
\end{equation}
\begin{equation}
C_4 = \left(2\alpha+\gamma_{\rm g}+\beta\gamma_{\rm c}\right)
      \left(\eta^2+\zeta^2+\frac{1}{4}\right)
    + \left[ \kappa_{\parallel} \eta^2 -
             \kappa_{\perp}\left(\frac{1}{2}+i\zeta\right)^2 \right]^2,
\end{equation}
\begin{equation}
C_3 = \left(4\alpha+2\gamma_{\rm g}+\beta\gamma_{\rm c}\right)
        \left(\eta^2+\zeta^2+\frac{1}{4}\right)
        \left[ \kappa_{\parallel} \eta^2 -
            \kappa_{\perp}\left(\frac{1}{2}+i\zeta\right)^2 \right]
    +\beta\kappa_{\perp}\left(\frac{1}{2}+i\zeta\right)
           \left[\eta^2-\left(\frac{1}{2}+i\zeta\right)^2 \right]
\end{equation}
\begin{eqnarray}
C_2 &=& \left[ 2\alpha(\gamma_{\rm g}+\beta\gamma_{\rm c})
              \left(\eta^2+\zeta^2+\frac{1}{4}\right) -
              (1+\alpha+\beta)
              \left(1+\alpha+\beta-\gamma_{\rm g}-\beta\gamma_{\rm c}\right)
        \right]\eta^2 
\nonumber \\
    &+&(2\alpha+\gamma_{\rm g})\left(\eta^2+\zeta^2+\frac{1}{4}\right)
        \left[ \kappa_{\parallel} \eta^2 -
               \kappa_{\perp}\left(\frac{1}{2}+i\zeta\right)^2 \right]^2
\nonumber \\
    &+& \beta\kappa_{\perp}\left(\frac{1}{2}+i\zeta\right)
        \left[\eta^2-\left(\frac{1}{2}+i\zeta\right)^2\right]
        \left[ \kappa_{\parallel}\eta^2
              -\kappa_{\perp}\left(\frac{1}{2}+i\zeta\right)^2 \right]
\end{eqnarray}
\begin{eqnarray}
C_1 &=& \left[ 2\alpha(2\gamma_{\rm g}+\beta\gamma_{\rm c})
             \left(\eta^2+\zeta^2+\frac{1}{4}\right)-
             (1+\alpha+\beta)(2+2\alpha+\beta-2\gamma_{\rm g}
             -\beta\gamma_{\rm c})\right]
\nonumber \\
    &\times& \left[ \kappa_{\parallel} \eta^2 -
             \kappa_{\perp}\left(\frac{1}{2}+i\zeta\right)^2 \right] \eta^2
     + \beta(1+\alpha+\beta)\eta^4(\kappa_{\perp}-\kappa_{\parallel})
\end{eqnarray}
\begin{eqnarray}
\label{eq:c0}
C_0 &=& \left[ 2\alpha\gamma_{\rm g}
             \left(\eta^2+\zeta^2+\frac{1}{4}\right)
             -(1+\alpha+\beta)(1+\alpha-\gamma_{\rm g})\right]
      \left[ \kappa_{\parallel} \eta^2 -
             \kappa_{\perp}\left(\frac{1}{2}+i\zeta\right)^2 \right]^2 \eta^2
\nonumber \\
    &+& \beta(1+\alpha+\beta)(\kappa_{\perp}-\kappa_{\parallel})
         \left[\kappa_{\parallel}\eta^2
        -\kappa_{\perp}\left(\frac{1}{2}+i\zeta\right)^2 \right] \eta^4.
\end{eqnarray}

\section{Results}

\subsection{Parameters}

The dispersion relation, equation~(\ref{eq:polynomial}) augmented by
equations~(\ref{eq:c5}-\ref{eq:c0}), gives
the linear growth rate, $n$, as a function of the azimuthal wave number,
$\eta$, and the vertical wavenumber, $\zeta$. It involves the parameters
$\alpha$, $\beta$, $\gamma_{\rm g}$, $\gamma_{\rm c}$, $\kappa_{\parallel}$,
and $\kappa_{\perp}$. In addition to $\gamma_{\rm g}=1$ and
$\gamma_{\rm c}=14/9$ which were specified in \S 2.1, $\alpha=1$ and $\beta=1$
will be used in calculation of the growth rate. Setting $\alpha=1$ and
$\beta=1$ means initially the magnetic and CR pressures are same as the gas
pressure. For the energy weighted CR diffusion coefficients,
$\kappa_{\parallel} = 3 \times 10^{28}$ cm$^2$ s$^{-1}$ and
$\kappa_{\perp} = 0.02 \kappa_{\parallel}$ will be taken as the fiducial
values (\S 2.2). Then, in units of $aH$, approximately
$(\kappa_{\parallel},\kappa_{\perp})=(100,2)$. Other values of
$(\kappa_{\parallel},\kappa_{\perp})$ will be also considered in
demonstrating the effect of CR diffusion.

\subsection{Parallel Diffusion
(Non-zero $\kappa_{\parallel}$ and $\kappa_{\perp} = 0$)}

\begin{figure}
\plotone{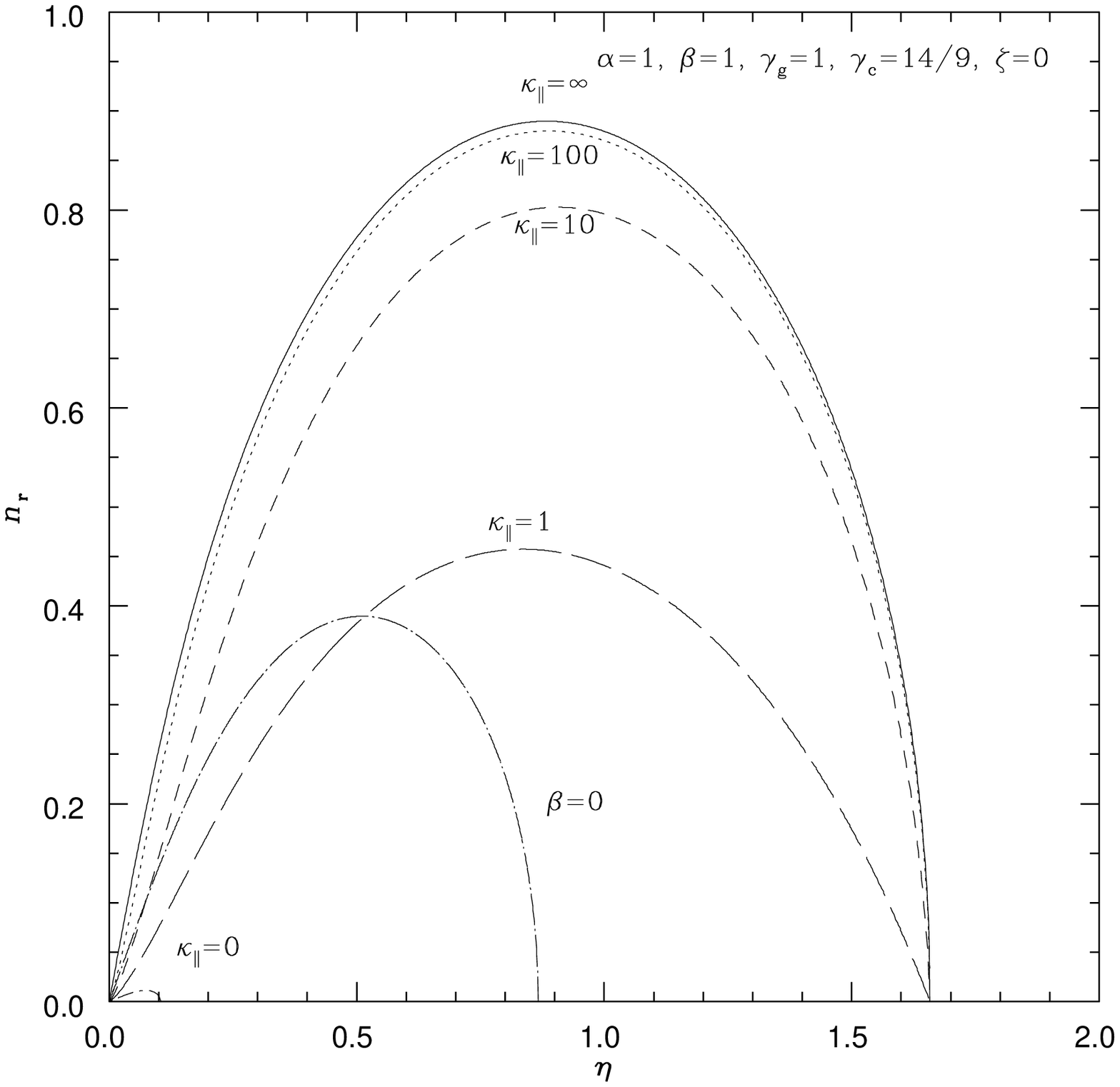}
\caption{Dispersion relations of the Parker instability with non-zero
$\kappa_{\parallel}$ and $\kappa_{\perp} = 0$. The growth rate (the largest
$n$) is presented as a function of the wavenumber along the initial magnetic
field direction. The vertical wavenumber along the direction of gravity was
set to be zero. The normalization units of time and length are
$2.4 \times 10^7$ yrs $(H/a)$ and 160 pc $(H)$, respectively. Each curve
is labeled by the value of $\kappa_{\parallel}$. Values of other parameters
are specified within the frame. The growth rate of the case without CRs
$(\beta=0)$ is also presented for comparison, where the same normalization
was applied.}
\end{figure}

We first check whether our dispersion relation recovers the ones
of previous works with simplified treatments of CR dynamics.
$d\delta P_c/dt=0$ of \citet{par66} is translated to $\gamma_{\rm c}=0$ and
$\kappa_{\parallel} = \kappa_{\perp} = 0$ in our formulation. The assumption
introduced by \citet{shu74} corresponds to
$\kappa_{\parallel} \rightarrow \infty$ with $\kappa_{\perp} = 0$. Although
they look different, both limits end with the same dispersion relation
\begin{equation}
\label{eq:4th}
n^4 + (2\alpha+\gamma_{\rm g})\left(\eta^2+\zeta^2+\frac{1}{4}\right)n^2 +
\left[ 2\alpha\gamma_{\rm g}\left(\eta^2+\zeta^2+\frac{1}{4}\right) -
(1+\alpha+\beta)\left(1+\alpha+\beta-\gamma_{\rm g}\right)\right]\eta^2 =0,
\end{equation}
which matches with those of \citet{par66} and \citet{shu74}.

We now study the case of non-zero parallel diffusion, but
no perpendicular diffusion. Note that with $\kappa_{\perp} = 0$, the initial
equilibrium is exact, and the ad hoc inclusion of a source is not necessary
(see \S 2.3). Hence, the dispersion relation is exact with its limit. Taking
$\kappa_{\perp}=0$, equations~(\ref{eq:polynomial})-(\ref{eq:c0}) reduce to
\begin{eqnarray}
\label{eq:5th}
&&
n^5 + \kappa_{\parallel} \eta^2 n^4 +   
(2\alpha+\gamma_{\rm g}+\beta\gamma_{\rm c})
\left(\eta^2+\zeta^2+\frac{1}{4}\right)n^3 +
(2\alpha+\gamma_{\rm g})\left(\eta^2+\zeta^2+\frac{1}{4}\right)
\kappa_{\parallel}\eta^2 n^2
\nonumber \\
&+&\left[2\alpha(\gamma_{\rm g}+\beta\gamma_{\rm c})
\left(\eta^2+\zeta^2+\frac{1}{4}\right) 
-(1+\alpha+\beta)(1+\alpha+\beta-\gamma_{\rm g}
-\beta\gamma_{\rm c})\right]\eta^2 n
\nonumber \\
&+& \left[2\alpha\gamma_{\rm g}\left(\eta^2+\zeta^2+\frac{1}{4}\right)
-(1+\alpha+\beta)(1+\alpha+\beta-\gamma_{\rm g})\right]
\kappa_{\parallel}\eta^4=0,
\end{eqnarray}
after factoring out $n+\kappa_{\parallel}\eta^2=0$ which represents
diffusive decay.

Figure 1 plots the growth rate (the largest $n$) as a function of
the azimuthal wavenumber, $\eta$, for  zero vertical wavenumber
$\zeta=0$ and several different values of $\kappa_{\parallel}$
including $\kappa_{\parallel}\rightarrow\infty$ and $\kappa_{\parallel}=0$.
Non-zero $\zeta$ reduces $n$. We note that the dispersion
relation (\ref{eq:5th}) is of fifth order and has five roots. As
$\kappa_{\parallel}\rightarrow\infty$, four reduce to the roots of
(\ref{eq:4th}), while the last one becomes $n+\kappa_{\parallel}\eta^2=0$
representing another mode of diffusive decay. Figure 1 also plots
the growth rate of the case without CRs $(\beta=0)$. The figure shows that
the growth rate increases with increasing $\kappa_{\parallel}$. Finiteness
of $\kappa_{\parallel}$ reduces the growth of the instability over that of
$\kappa_{\parallel}\rightarrow\infty$, because there is a gradient of CR
pressure along magnetic field lines. The gradient hinders the falling motion
of gas from arc regions to valleys and slows down the development of
the instability. However, with the value of $\kappa_{\parallel}$ expected
in the ISM, $\sim100$, the growth rate is close to that of
$\kappa_{\parallel}\rightarrow\infty$. So we conclude that
$d\delta P_c/dt=0$ or $\kappa_{\parallel} \rightarrow \infty$ with
$\kappa_{\perp} = 0$ used in previous analyses were good approximations
and produced quantitatively correct results. With $\kappa_{\parallel}=100$,
the maximum growth rate and the critical wavenumber where $n \rightarrow 0$
are about twice larger than those of $\beta=0$, indicating CRs can enhance
the instability significantly.

One thing to note is that the critical wavenumber is independent of
$\kappa_{\parallel}$, once $\kappa_{\parallel} > 0$, and is given as
\begin{equation}
\eta_{\rm c}^2 = 
(1+\alpha+\beta)(1+\alpha+\beta-\gamma_{\rm g}) / (2\alpha\gamma_{\rm g})
-\zeta^2 - \frac{1}{4}. 
\end{equation}
The reason is the following. It takes infinitely long for such a marginally
stable state to be developed, because of the zero growth rate. This means
that, however small the diffusion is, the system has enough
time to diffuse across any gradient of CR pressure along magnetic field lines.
However, with no diffusion, $\kappa_{\parallel} = 0$, the critical wavenumber
is different and given as
\begin{equation}
\eta_{\rm c}^2 = 
(1+\alpha+\beta)(1+\alpha+\beta-\gamma_{\rm g}-\beta\gamma_{\rm c}) /
[2\alpha(\gamma_{\rm g}+\beta\gamma_{\rm c})] -\zeta^2 - \frac{1}{4}. 
\end{equation}
which is much smaller for the parameters we employed. And as the matter of
fact, the growth rate is much smaller too, and even smaller than that of
$\beta=0$. This is because without diffusion, CRs accumulate at valleys
along with gas, and their pressure pushes gas out of valleys exerting
a stabilizing effect.

\subsection{Perpendicular Diffusion 
($\kappa_{\parallel} = 0$ and Non-zero $\kappa_{\perp}$)}

\begin{figure}
\plotone{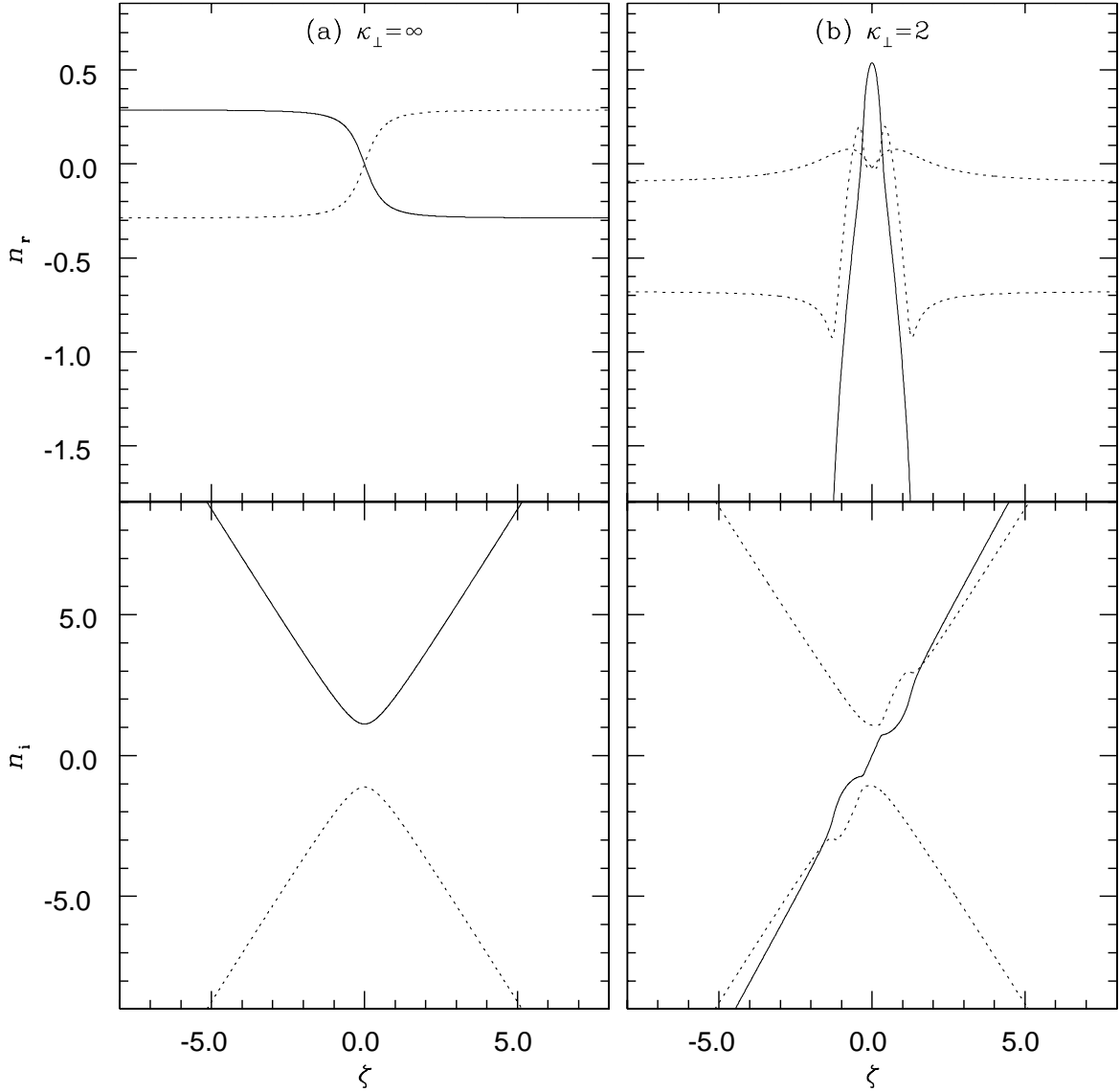}
\caption{Dispersion relations with $\kappa_{\parallel}=0$ and non-zero
$\kappa_{\perp}$ for zero azimuthal wavenumber, $\eta=0$. The real part,
$n_r$, and imaginary part, $n_i$, of the grow rate are presented as a
function of the vertical wavenumber, $\zeta$, for two different values
of $\kappa_{\perp}$'s. The normalization units of time and length are
$2.4 \times 10^7$ yrs $(H/a)$ and 160 pc $(H)$, respectively. Values
of other parameters are $\alpha=1, \beta=1, \gamma_{\rm g}=1$, and
$\gamma_{\rm c}=14/9$.}
\end{figure}

We start to investigate the effect of non-zero perpendicular
diffusion on the Parker instability, by looking at the case of
$\kappa_{\parallel}=\eta=0$. This case describes the acoustic instability
of CR mediated gas, which was studied by \citet{dru86} and \citet{kan92},
but with magnetic field. From the full dispersion
relation~(\ref{eq:polynomial})-(\ref{eq:c0}), we get in this case
\begin{eqnarray}
\label{eq:3th}
\left[n-\kappa_{\perp}\left(\frac{1}{2}+i\zeta\right)^2\right]
\Bigg\{n^3 &-& \kappa_{\perp}\left(\frac{1}{2}+i\zeta\right)^2 n^2
+(2\alpha+\gamma_{\rm g}+\beta\gamma_{\rm c})
\left(\frac{1}{4}+\zeta^2\right)n\nonumber \\
&-&\kappa_{\perp} \left[\beta+(2\alpha+\gamma_{\rm g})
\left(\frac{1}{2}-i\zeta\right)\right]
\left(\frac{1}{2}+i\zeta\right)^3\Bigg\} = 0.
\end{eqnarray}
In the limit of zero diffusion, the dispersion relation,
$n = \pm i \sqrt{(2\alpha+\gamma_{\rm g}+\beta\gamma_{\rm c})(1/4+\zeta^2)}$,
describes a pair of magnetosonic waves propagating upwards and downwards.
The term of $\beta\gamma_{\rm c}$ represents the effect of additional
support of CR pressure on top of the magnetosonic waves. The $1/4$
term comes from the $\exp(z/2)$ factor in the normal mode of perturbations
in equation~(\ref{eq:perturbation}), added to account the stratified
background. As a matter of fact, all the $1/2$ and $1/4$ terms in
the dispersion relation~(\ref{eq:3th}) come from the same origin.

Figure 2 plots the growth rate as a function of the vertical wavenumber,
$\zeta$, for two non-zero $\kappa_{\perp}$'s. For other parameters,
the values listed in \S 3.1 were used. As pointed out in \citet{dru86} and
\citet{kan92}, due to the ad hoc source term, which is necessary to sustain
the initial equilibrium state, the analysis is justified rigorously only in
the limit of $k_z \gg 1/H$ or $\zeta \gg 1$ in our normalized units. In that
limit, the growth rate of non-trivial modes is
\begin{equation}
\label{eq:infinitezeta}
n = \pm i \sqrt{2 \alpha + \gamma_{\rm g}}~\zeta
- \frac{\beta\gamma_{\rm c}}{2\kappa_{\perp}}
\mp \frac{\beta}{2\sqrt{2 \alpha + \gamma_{\rm g}}}.
\end{equation}
The first term on the right hand side represents magnetosonic waves,
which are shown in the bottom panels of Figure 2. With non-zero
diffusion, the CR perturbation associated with with small wavelengths is
wiped out due to the diffusive nature. So CR pressure $(\beta)$ does not
contribute to the speed of the waves. The second and third terms are
attributed to the acoustic instability, if their sum is positive, or
if $\kappa_{\perp}$ is sufficiently large, which is shown in the upper
left panel of Figure 2. The above limiting growth rate matches exactly
with that of \citet{kan92}, if the magnetosonic waves are replaced by sound
waves.

The physical mechanism of the acoustic instability is the following.
Perturbations in a gas mediated by CRs with a gradient can become unstable,
because the CR pressure perturbation is reduced by diffusion. In the limit of
large diffusion, CRs are completely decoupled from the gas for small scale
perturbations but the gradient of CR pressure will remain the same. In this
limit, let us suppose a constant volume force, $F$, is exerted on the gas
so that the acceleration is $F/\rho$. Then, with $- F{\delta\rho/\rho^2}$,
compressed regions in a wave train will be accelerated in the opposite
direction of the applied force,
while decompressed regions will be accelerated in
the direction of the force. As a result, oscillating density disturbances
(or magnetosonic/sound waves) moving opposite to the direction of the force
will suffer an extra restoring force and their amplitude will grow.
On the other hand, waves propagating in the other direction will be
damped. This explains why disturbances traveling in one direction will
be amplified, while those traveling in the opposite direction will decay.
The instability works only if  the following two conditions are satisfied:
1) the perturbation wavelength is shorter than the scale height of CR
pressure, or $\eta > 1$, and 2) the scale height of CR pressure is smaller
than the diffusion length associated with sound speed, or $\kappa_{\perp}>1$,
in our normalized units.

In addition to the acoustic instability (upper left panel), Figure 2
shows the unstable nature in the regime of $\kappa_{\perp}\sim 1$ and
$\zeta \la 1$, especially in $\zeta = 0$ which corresponds to the state
of no perturbation at all (upper right panel). Without any viable
mechanism, we attribute this to the artifact of the ad hoc source
term along with the $\exp(z/2)$ factor in the normal mode. The detailed
analysis shows that it is dominated by the perturbation of CRs,
confirming that it is not related to the acoustic instability.

\subsection{Non-zero Parallel and Perpendicular Diffusions}

\begin{figure}
\plotone{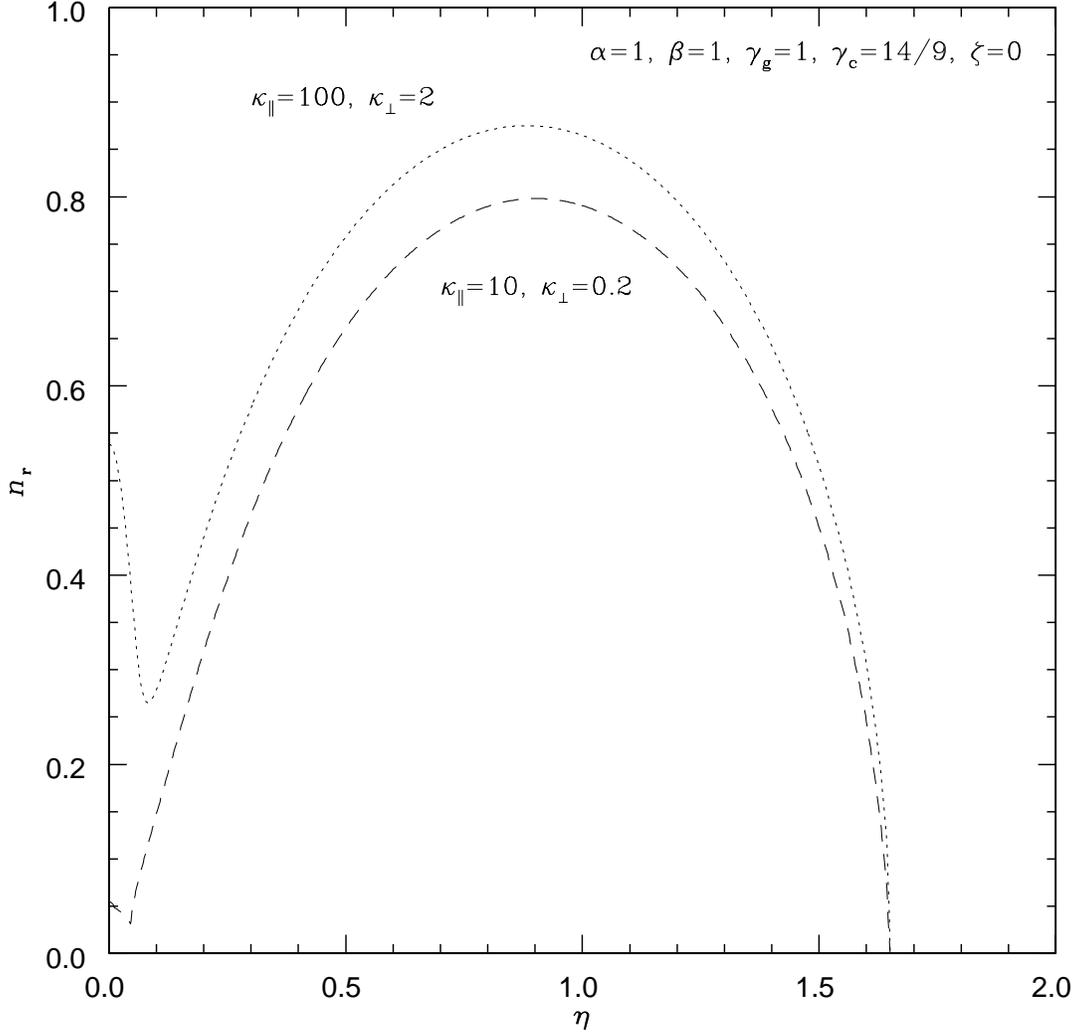}
\caption{Dispersion relations of the Parker instability with non-zero
$\kappa_{\parallel}$ and non-zero $\kappa_{\perp}$. The growth rate (the
largest $n$) is presented as a function of the wavenumber along the initial
magnetic field direction. The vertical wavenumber along the direction of
gravity was set to be zero. The normalization units of time and length are
$2.4 \times 10^7$ yrs $(H/a)$ and 160 pc $(H)$, respectively. Each curve
is labeled by the values of $\kappa$'s. Values of other parameters
are specified within the frame.}
\end{figure}

Finally, we consider the effect of non-zero perpendicular diffusion on
top of non-zero parallel diffusion. Figure 3 plots, from the
dispersion relation~(\ref{eq:polynomial})-(\ref{eq:c0}), the growth rate
(the largest $n$) as a function of the azimuthal wavenumber, $\eta$,
for two different sets of the values of $\kappa$'s and zero vertical
wavenumber $\zeta=0$. Comparing Figure 1 and Figure 3, it can be seen that
although $\kappa_{\perp}$ is expected to reduce $n$, the growth rates with
the same $\kappa_{\parallel}$ are almost identical except around $\eta=0$.
This is because $\kappa_{\perp} \ll \kappa_{\parallel}$ in
the ISM. The peak at $\zeta=0$ is again attributed to the artifact of
the ad hoc source term along with the $\exp(z/2)$ factor in the normal
mode, as explained in the previous section. So we conclude that with realistic
values of CR diffusion in the ISM, the effect of non-zero perpendicular
diffusion is mostly negligible.

\section{SUMMARY AND DISCUSSIONS}

The Parker instability in the ISM is induced by the buoyancy of magnetic
field as well as CRs. Hence, its analysis can be completed with full
treatment of CR dynamics. However, previous analyses incorporated CRs with
simplified assumptions or ignored CRs completely, partially due to the lack
of available CR physics. For instance, although diffusion of CRs
is finite in the ISM, \citet{par66} and \citet{shu74} assumed that the
diffusion along magnetic field lines is large enough that there is no CR
pressure gradient along them, while the diffusion across field lines
is neglected. In this contribution, we have relaxed this assumption and
studied the role of full CR dynamics with finite CR diffusion in the Parker
instability. For it, first, the values of the energy weighted mean diffusion
coefficients, which are applicable to the scales relevant to the Parker
instability, have been estimated as
$\kappa_{\parallel} \simeq 3 \times 10^{28}$ cm$^2$ s$^{-1}$ and
$\kappa_{\perp} = 0.02 \kappa_{\parallel}$. Then, a standard normal mode
analysis has been performed in the two-dimensional plane defined by the
gravity and the initial magnetic field. Linearized perturbation equations
have been combined into the dispersion relation~(\ref{eq:polynomial}) of
a polynomial of 6th order in the growth rate $n$ with complex coefficients
given by~(\ref{eq:c5})-(\ref{eq:c0}).

It has been shown that finiteness of parallel diffusion slows down
the development of the Parker instability. However, with
$\kappa_{\parallel} = 3 \times 10^{28}$ cm$^2$ s$^{-1}$ in the ISM,
the maximum growth rate is smaller only by a couple of percents than that for
$\kappa_{\parallel} \rightarrow \infty$, and the range of unstable
wavenumbers remains the same. Inclusion of perpendicular diffusion with
$\kappa_{\perp} = 0.02 \kappa_{\parallel}$ doesn't change the growth rate
noticeably. That is, the original Parker's approximation of infinite
$\kappa_{\parallel}$ and $\kappa_{\perp}=0$ was a good one and produced
a quantitatively correct result. Hence, we conclude that CRs can enhance
the Parker instability significantly, by increasing the maximum growth rate
by a factor of up to two or so. We would like to note that this result
disagrees with that of \citet{nel85}, who found that CRs could stabilize,
rather than destabilize, the Parker instability.

As noted in the introduction, recent studies of the Parker instability,
where the random component as well as the regular component of magnetic
field were considered \citep{par00,kim01}, showed that the random component
of strength comparable to that of regular component
($\delta B^2/B_0^2 \ga 0.5$) can stabilize the instability
completely. For smaller $\delta B^2/B_0^2$, the instability is still
operating, but with reduced growth rate and vanishing wavenumber along
the radial direction of the Galaxy. Hence, it was concluded that the
Parker instability alone has a difficult time forming Galactic structures
like giant molecular clouds. However, our new result suggests that
the Parker instability might be preserved once CRs are incorporated, since
inclusion of CRs increases not only the growth rate and but also the
range of unstable wavenumbers. Settling this issue would require full
three-dimensional analysis with random component of magnetic field,
which we leave for future work.

Finally, we comment on the applicability of an analysis that
assumes a smooth distribution of CRs, even though they are thought
to be products of discrete sources; namely, of supernova remnants.
For the purposes of the present calculation the approximation should
be quite adequate, in fact.
Observations of secondary pion-produced
$\gamma$-rays show that the galactic hadronic CR distribution is 
smooth on large scales \citep[see, e.g.,][]{bsbc86}. 
This is very reasonable in light of the long containment time of
such CRs in the galaxy \citep[$\sim 10^7$ yrs; see, e.g.,][]{conn98}
and their associated diffusion and advection. 
Recent, sophisticated models of the CR
distribution including stochastic sources along with numerous
experimental constraints lead above a few hundred MeV to very smooth 
distributions outside the CR acceleration sites \citep{sm01}.
Further, there are good arguments that
small, isolated supernova remnants expanding into dense media
could be far less common and less important sources of CRs than 
supernova remnants inside large, 
low density bubbles where the freshly accelerated CRs would be 
more broadly distributed \citep[see, e.g.,][]{hlr98}.

\acknowledgments

The work was supported in part by KRF through grant KRF-2000-015-DS0046.
We thank the anonymous referee for constructive comments.


\begin{thebibliography}{}

\bibitem[Beck et al.(1996)]{bbmss96}
Beck, R., Brandenburg, A., Moss, D., Shukurov, A., \& Sokoloff, D.
1996, \araa, 34, 155

\bibitem[Bieber \& Matthaeus(1997)]{bm97}
Bieber, J. W., \& Matthaeus, W. H.
1997, \apj, 485, 655

\bibitem[Blandford \& Eichler(1987)]{be87}
Blandford, R. D., \& Eichler, D.
1987, \physrep, 154, 1

\bibitem[Bloeman et al.(1986)]{bsbc86}
Bloeman, J., Strong, A. W., Blitz, L., Cohen, R. S., Dame, T. M., Grabelsky,
D. A., Hermsen, W., Lebrun, F., Mayer-Hasselwander, H. A., \& Thaddeus, P.
1986, \aap, 154, 25

\bibitem[Casse et al.(2002)]{clp02}
Casse, F., Lemoine, M., \& Pelletier, G.
2002, \prd, 65, 023002

\bibitem[Connell(1998)]{conn98}
Connell, J. J.
1998, \apjl, 501, L59

\bibitem[Drury \& V\"olk(1981)]{dru81}
Drury, L. O'C, \& V\"olk, H. J.
1981, \mnras, 248, 344

\bibitem[Drury \& Falle(1986)]{dru86}
Drury, L. O'C, \& Falle, S. A. E. G.
1986, \mnras, 223, 353

\bibitem[Drury(1983)]{dru83}
Drury, L. O'C.
1983, Rept. Prog. Phys., 46, 973

\bibitem[Falgarone \& Lequeux(1973)]{fl73}
Falgarone, E., \& Lequeux, J.
1973, \aap, 25, 253

\bibitem[Giacalone \& Jokipii(1999)]{gia99}
Giacalone, J., \& Jokipii, J. R.
1999, \apj, 520, 204

\bibitem[Giz \& Shu(1993)]{giz93}
Giz, A. T., \& Shu, F. H.
1993, \apj, 404, 185

\bibitem[Hanasz \& Lesch(2000)]{han00}
Hanasz, M., \& Lesch, H.
2000, \apj, 543, 235

\bibitem[Higdon et al.(1998)]{hlr98}
Higdon, J. C., Lingenfelter, R. E. \& Ramaty, R.
1998, \apjl, 509, L33

\bibitem[Jones \& Kang(1990)]{jon90}
Jones, T. W., \& Kang, H.
1990, \apj, 363, 499

\bibitem[Kang et al.(1992)]{kan92}
Kang, H., Jones, T. W., \& Ryu, D.
1992, \apj, 385, 193

\bibitem[Kim et al.(1997)]{kim97}
Kim, J., Hong, S. S., \& Ryu, D.
1997, \apj, 485, 228

\bibitem[Kim \& Hong(1998)]{kim98}
Kim, J., \& Hong, S. S.
1998, \apj, 507, 254

\bibitem[Kim et al.(2000)]{kim00}
Kim, J., Franco, J., Hong, S. S., Santill\'an, A., \& Martos, M. A.
2000, \apj, 531, 873

\bibitem[Kim \& Ryu(2001)]{kim01}
Kim, J., \& Ryu, D.
2001, \apj, 561, L135

\bibitem[Kuznetsov \& Ptuskin(1983)]{kp83}
Kuznetsov, V. D., \& Ptuskin, V. S.,
1983, Sov. Astron. Lett., 9, 75

\bibitem[Nelson(1985)]{nel85}
Nelson, A. H.
1985, \mnras, 215, 161

\bibitem[Parker(1966)]{par66}
Parker, E. N.
1966, \apj, 145, 811

\bibitem[Parker(1967)]{par67}
Parker, E. N.
1967, \apj, 149, 535

\bibitem[Parker \& Jopikii(2000)]{par00}
Parker, E. N. \& Jopikii, J. R.
2000, \apj, 536, 334

\bibitem[Santill\'an et al.(2000)]{san00}
Santill\'an, A., Kim, J., Franco, J., Martos, M. A., Hong, S. S., \& Ryu, D.
2000, \apj, 545, 353

\bibitem[Shu(1974)]{shu74}
Shu, F. H.
1974, \aap, 33, 55

\bibitem[Skilling(1975)]{ski75}
Skilling, J.
1975, \mnras, 172, 557

\bibitem[Strong \& Moskalenko(1998)]{sm98}
Strong, A. W., \& Moskalenko, I. V.
1998, \apj, 509, 212

\bibitem[Strong \& Moskalenko(2001)]{sm01}
Strong, A. W. \& Moskalenko, I. V.
2001, Proceedings of the 27th ICRC (Hamburg), p 1942

\bibitem[Zweibel \& Heiles(1997)]{zh97}
Zweibel, E. G. \& Heiles, C.
1997, \nat, 385, 131

\end{thebibliography}
\end{document}